# A Personalized Web Page Content Filtering Model Based on Segmentation

K.S.Kuppusamy[1] and G.Aghila[2]

[1]Department of Computer Science, School of Engineering and Technology, Pondicherry University, Pondicherry, India
`kskuppu@gmail.com`
[2]Department of Computer Science, School of Engineering and Technology, Pondicherry University, Pondicherry, India
`aghilaa@yahoo.com`

*ABSTRACT*

*In the view of massive content explosion in World Wide Web through diverse sources, it has become mandatory to have content filtering tools. The filtering of contents of the web pages holds greater significance in cases of access by minor-age people. The traditional web page blocking systems goes by the Boolean methodology of either displaying the full page or blocking it completely. With the increased dynamism in the web pages, it has become a common phenomenon that different portions of the web page holds different types of content at different time instances. This paper proposes a model to block the contents at a fine-grained level i.e. instead of completely blocking the page it would be efficient to block only those segments which holds the contents to be blocked. The advantages of this method over the traditional methods are fine-graining level of blocking and automatic identification of portions of the page to be blocked. The experiments conducted on the proposed model indicate 88% of accuracy in filtering out the segments.*

## 1. INTRODUCTION

The World Wide Web (WWW) has become the biggest repository of information known to the mankind. Yet another aspect which makes the World Wide Web more powerful is the ease with which this largest repository can be accessed. With the prolific improvements in the Web Search Engine's functionalities, the distance to any information available in the World Wide Web is "a single click".

Though it can be considered an advantage, it poses certain vulnerabilities as well. With the floodgates of information on World Wide Web open, the exposure to diverse information to all users has created the necessity for the some sort of filtering mechanisms to this information consumption process.

This paper proposes a technique which would facilitate this filtering mechanism. The objectives of this paper are as listed below:

- Proposing a model for web page content filtering based on segmentation.
- Incorporation of personalization in the proposed model to enhance the web content filtering process.

The remainder of this paper is organized as follows: In Section 2, some of the related works carried out in this domain are explored. Section 3 deals with the proposed model's mathematical representation and algorithms. Section 4 is about prototype implementation and experiments. Section 5 focuses on the conclusions and future directions for this research work.

## 2. RELATED WORKS

This section would highlight the related works that have been carried out in this domain. The proposed model incorporates the following two major fields of study:

- The Web Content Filtering
- Web Page Segmentation

### 2.1 The Web Content Filtering

Content Filtering Systems for web pages is an active research topic primarily due to following reasons: It protects users (especially minor-age people) from unwanted content; the resources on the network can be saved from unwanted usage like playing network games in an office network etc. There exist many approaches to Content Filtering Systems. Some of them are as listed below:

- Rating Systems
- Black Listing / White Listing
- Keyword blocking
- 

In Rating Systems users are asked to rate a web site for its content. This rating would be used as a tool for filtering [1]. The black listing / white listing maintains a set of URLs manually prepared for filtering. The problem with this approach is the scalability. There exist many tools available to perform content filtering using above specified methods [2], [3], [4].

The text classification based approach is explored in [5], [6]. The approach that has been chosen to facilitate filtering in this paper is a variation of keyword based blocking method.

### 2.2 Web Page Segmentation

Web page segmentation is an active research topic in the information retrieval domain in which a wide range of experiments are conducted. Web page segmentation is the process of dividing a web page into smaller units based on various criteria. The following are four basic types of web page segmentation method:

- Fixed length page segmentation
- DOM based page segmentation
- Vision based page segmentation
- Combined / Hybrid method

A comparative study among all these four types of segmentation is illustrated in [7]. Each of above mentioned segmentation methods have been studied in detail in the literature. Fixed length page segmentation is simple and less complex in terms of implementation but the major problem

International Journal of Information Sciences and Techniques (IJIST) Vol.2, No.1, January 2012

with this approach is that it doesn't consider any semantics of the page while segmenting. In DOM base page segmentation, the HTML tag tree's Document Object Model would be used while segmenting. An arbitrary passages based approach is given in [8]. Vision based page segmentation (VIPS) is in parallel lines with the way, humans views a page. VIPS [9] is a popular segmentation algorithm which segments a page based on various visual features.

Apart from the above mentioned segmentation methods a few novel approaches have been evolved during the last few years. An image processing based segmentation approach is illustrated in [10]. The segmentation process based text density of the contents is explained in [11]. The graph theory based approach to segmentation is presented in [12]. Repetition-based web page segmentation by detecting tag patterns for small-Screen Devices is explored in [13]. One of the approaches for web page segmentation for specific domains is detailed in [14]. A tree clustering based segmentation approach is provided in [15].

## 3. THE MODEL

This section elaborates about the mathematical model of the proposed system. The corresponding algorithm to carry out the task specified in the model is also explored in this section. The block diagram of the proposed model is as shown in Figure 1. It contains the following components:

**Page Segmentor**: This component is responsible for segmenting the contents of the page in to logically relevant units.

**Personalizer**: This component handles the personalization of filtering. The Personalizer holds the profile-bag which contains user preferences.

**Segment Filter**: Segment filter is another component in the model which handles individual segments and decides whether this segment should be incorporated in the filtered page or not.

### 3.1 Mathematical Model

In the proposed model each page that the user requests need to be segmented for filtration. Let us denote the source page by $\Phi$. The source page $\Phi$ has to be segmented in to various logically coherent parts.

The source page $\Phi$ would be mapped as a DOM (Document Object Model) tree. The individual nodes of the DOM tree are processed by parsing the tree. The "block level" and "non-block level" nodes are identified and they are used as the building block of the individual segments.

The approach followed in this paper also incorporates the densitometry concepts in the segment building process. The densitometry considers the density of text present at a block unit in performing the segmentation process. As a result of the above mentioned process, the source page $\Phi$ is segmented in to various units as shown in (1).

$$\Phi = \{\mu_1, \mu_2, \mu_3 ... \mu_n\} \quad (1)$$

The segmentation process shown in (1) is performed by the "Page Segmentor" component in the proposed model.

After the completion of the segmentation process, each of the segments needs to be processed individually. This is performed by the "Segment Filter Component". Each segment $\mu_i$ is represented as a triple containing three components as shown listed below:

- Text
- Link
- Image

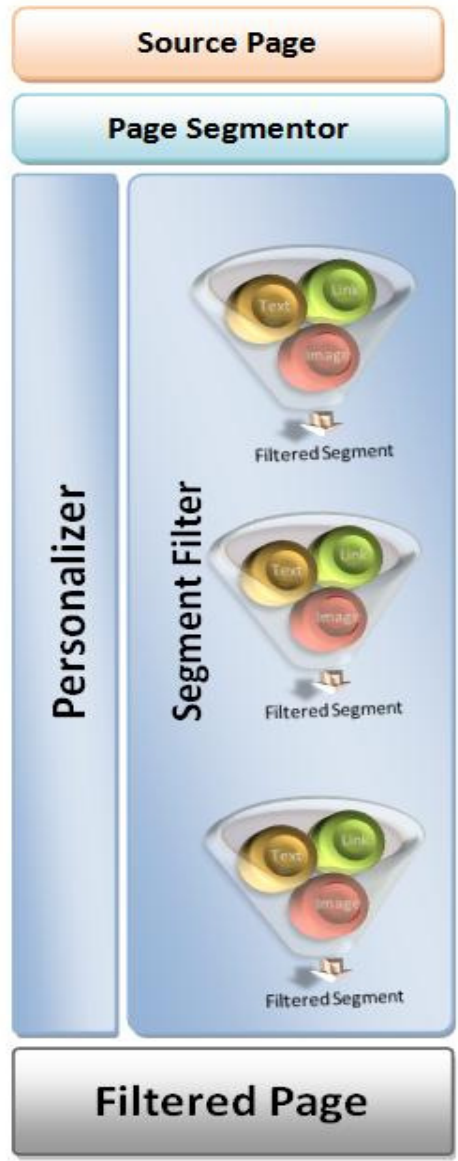

Figure 1. Block Diagram of the Model

The segment triple containing text, link and image is represented as shown in (2).

$$\mu_i = \{\Psi, \Lambda, \Theta\} \quad (2)$$

The triple $\mu_i$ can be expanded as shown in (3).

International Journal of Information Sciences and Techniques (IJIST) Vol.2, No.1, January 2012$$\mu_i = \begin{Bmatrix} [\eta_1, \eta_2, \ldots \eta_p] & \forall \eta_i \in \Psi \\ [\kappa_1, \kappa_2, \ldots \kappa_q] & \forall \kappa_i \in \Lambda \\ [\lambda_1, \lambda_2, \ldots \lambda_r] & \forall \lambda_i \in \Theta \end{Bmatrix} \qquad (3)$$

In (3) $[\eta_1, \eta_2, \ldots \eta_p]$ represent the text elements present in the segment under consideration; $[\kappa_1, \kappa_2, \ldots \kappa_q]$ represent the individual links presents in the segment and $[\lambda_1, \lambda_2 \ldots \lambda_r]$ represent the image elements present in the segments.

The individual segments need to be processed for each of these three components to decide whether this segment can be allowed for display or it needs to be blocked. In order to perform this, segment filter component includes three sub-components a) Text Filter, b) Link Filter and c) Image Filter. The focus of this research work is on the effect of segmentation and personalization. The actual filtration process can be either simple keyword based or it can be customized according to the requirements of implementation.

The proposed model incorporates personalization aspect. The user can configure the filter according to his/her requirements. The user preferences are represented using "Profile Bag". The profile bag involves two different tracks. These tracks are "Like Track" and "Un-Like Track".

The block diagram of profile-bag is as shown in Figure 2. The figure consists of three horizontal layers. The top layer denotes the overall profile-bag. The middle one represents the "Like-Track" and "Un-Like Track". The bottom layer in the Figure 2 denotes the keywords which form the "Like-Track" and "Un-Like Track".

The profile bag is represented in the model as $\Gamma$. The two different tracks of $\Gamma$ are represented as shown in (4).

$$\Gamma = \langle \omega/\sigma \rangle \qquad (4)$$

In (4) $\omega$ represent the "Like Track" and $\sigma$ represent the "Un-Like Track" of the profile bag. Both $\omega$ and $\sigma$ contains keywords that represent the user preferences. The keywords in $\omega$ adds a positive booster and the keywords in $\sigma$ adds a negative booster.

The filtration process can be represented as shown in (5). As a result of (5) the Text Weight, Link Weight and Image Weight are calculated as the sum of number of terms common between $\omega$ and elements giving a "+1" weight and number of terms common between $\sigma$ and elements giving a "-1" weight.

45

$$\{\Psi, \Lambda, \Theta\} = \left\{ \begin{array}{ll} \left|\dfrac{[\eta_1, \eta_2, \ldots \eta_p]}{\Gamma}\right| & \forall \eta_i \in \Psi \\ \left|\dfrac{[\kappa_1, \kappa_2, \ldots \kappa_q]}{\Gamma}\right| & \forall \kappa_i \in \Lambda \\ \left|\dfrac{[\lambda_1, \lambda_2, \ldots \lambda_r]}{\Gamma}\right| & \forall \lambda_i \in \Theta \end{array} \right\}$$

(5)

If the sum of weights of all these three components exceeds a threshold level the segment is displayed otherwise it is blocked.

$$\underline{\Phi} = \left\{ \begin{array}{ll} \forall \mu_i \in \Phi : if\left(|\Psi| + |\Lambda| + |\Theta|\right) \geq \delta & \underline{\Phi} \cup \mu_i \\ else & \underline{\Phi} \cup \mu_z \end{array} \right\}$$

(6)

In (6), $\underline{\Phi}$ represents the filtered page in which segments whose weight has been calculated above the threshold limit are incorporated. When the weight is less than the threshold then a dummy segment $\mu_z$ holding the message "segment blocked" would be added to the page.

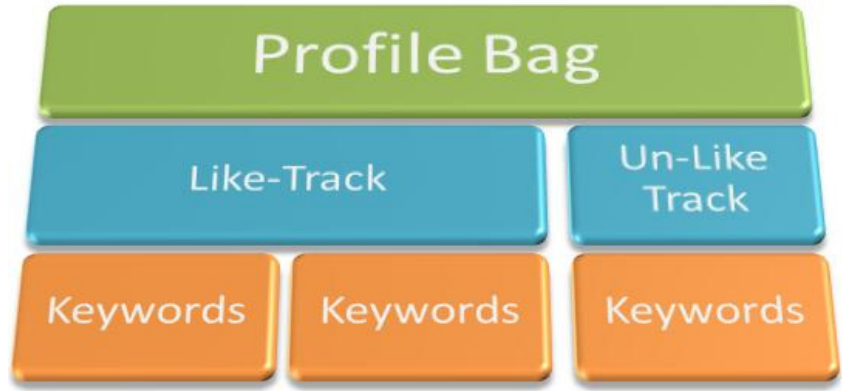

Figure 2. The User Profile - Bag

The dummy segment which would replace the filtered segment can be custom defined. The proposed model has another feature called "link hiding". In the case of link hiding, if the content to be blocked is having a hyperlink, instead of removing the content, the hyperlink alone can be removed which creates the similar impact as removing the content.

### 3.2 The Algorithm

The algorithmic representation of the steps involved in the above explained model is explored in this section.

```
Algorithm SegmentFilter
Input: Source Web Page $\Phi$ , profile bag $\Gamma$

Output : Filtered Page $\underline{\Phi}$

Begin
        Segment the source page using page segmentor
        $\Phi = \{\mu_1, \mu_2, \mu_3 ... \mu_n\}$

        Initialize $\underline{\Phi}$ to NULL
        For each segment $\mu_i$
        begin
                Parse the segment $\mu_1$ into components $\{\Psi, \Lambda, \Theta\}$
                Calculate Text weight $|\Psi|$ = TF ($\Psi / \Gamma$)
                Calculate Link Weight $|\Lambda|$ = LF ($\Lambda / \Gamma$)
                Calculate Image Weight $|\Theta|$ = IF ($\Theta / \Gamma$)
                If $(|\Psi| + |\Lambda| + |\Theta|) \geq \delta$ then
                        $\underline{\Phi} = \underline{\Phi} \cup \mu_i$
                Else
                        $\underline{\Phi} = \underline{\Phi} \cup \mu_z$
                End
                Return ($\underline{\Phi}$)
End
```

In the algorithm, TF, LF and IF refers to text filter, link filter and image filter respectively.

## 4. EXPERIMENTS AND RESULT ANALYSIS

The proposed model has been implemented as prototype for experimentation. The prototype implementation is done with the software stack including Linux, Apache, MySql and PHP. For client side scripting JavaScript is used. With respect to the hardware, a Core i3 processor system with 3 GHz of speed, 8 GB of RAM is used. The internet connection used in the experimental setup is a 128 Mbps leased line.

The screenshots of the prototype implementation are as shown in the Figure 3 and Figure 4. The screenshot shown in Figure 3 is of the original source page.

International Journal of Information Sciences and Techniques (IJIST) Vol.2, No.1, January 2012

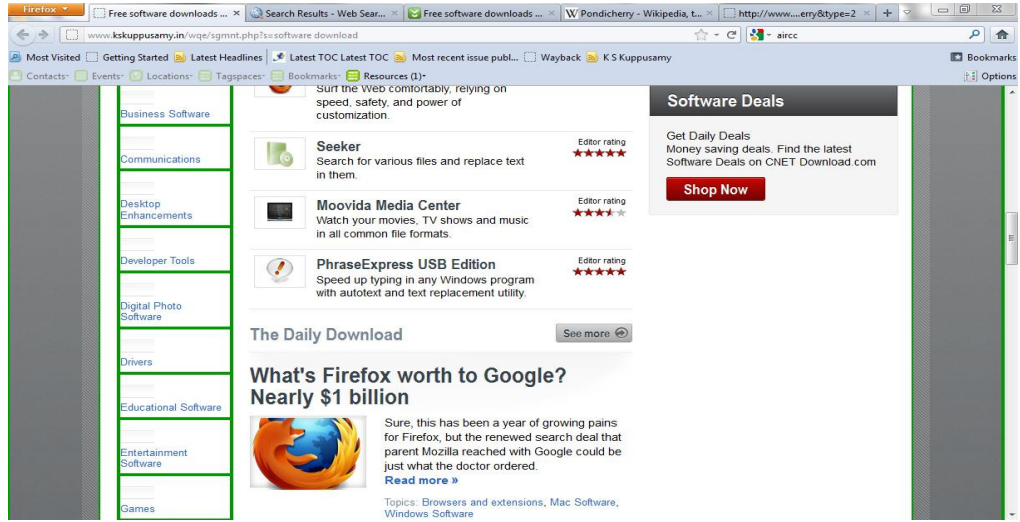

Figure 3. The Source Page

The page segments are filtered out based on the filtering preferences set up. The resultant page is as shown in Figure 4.

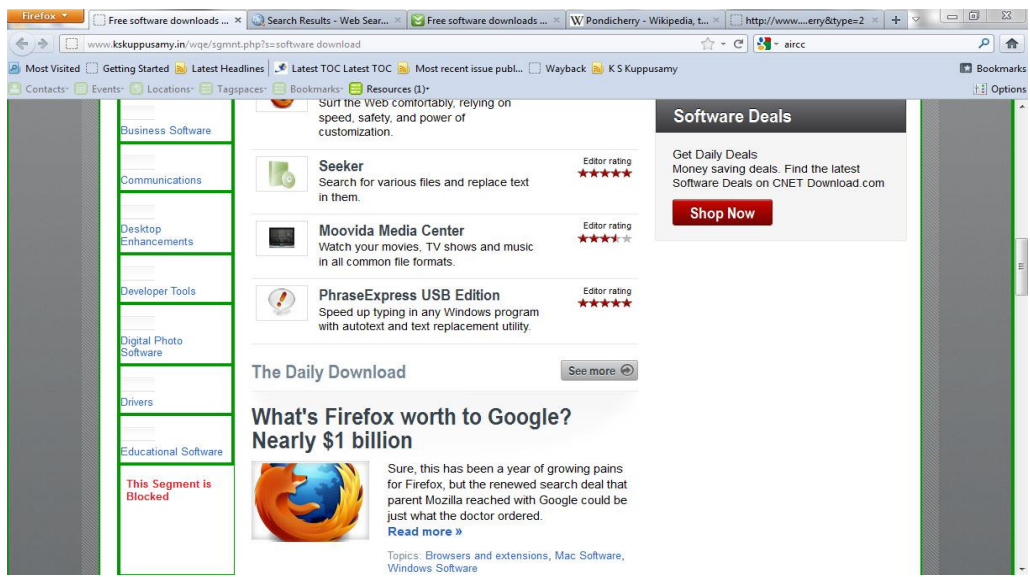

Figure 4. The Page after filtering the unwanted segments.

In Figure 4, it can be noted that the segments containing the terms "Entertainment Software" and "Games" are filtered out as per the filtering preferences set. The contents of Table 1 list out the experimental results conducted on the proposed content filtering model. In the Table 1, MSC indicates the mean segment count, MFSC stands mean filtered segment count, MFP is mean false positives and MFN is mean false negative.

Table 1. Experimental Results of the proposed model

| Session ID | MSC | MFSC | MFP | MFN | Accuracy (%) |
|---|---|---|---|---|---|
| 1 | 27.52 | 5.2 | 1.2 | 1.5 | 90.189 |
| 2 | 30.25 | 3.5 | 0.8 | 1.2 | 93.388 |
| 3 | 43.53 | 4.5 | 1.3 | 1.3 | 94.027 |
| 4 | 20.67 | 2.7 | 0.7 | 0.2 | 95.646 |
| 5 | 18.45 | 1.5 | 1.6 | 0.4 | 89.16 |
| 6 | 14.66 | 2.3 | 3.5 | 0.5 | 72.715 |
| 7 | 16.78 | 4.3 | 3.1 | 1.1 | 74.97 |
| 8 | 17.67 | 1.3 | 1.2 | 1.5 | 84.72 |
| 9 | 14.85 | 1.8 | 0.5 | 0.8 | 91.246 |
| 10 | 25.52 | 2.6 | 0.9 | 0.9 | 92.947 |
| 11 | 12.45 | 5.2 | 0.9 | 1.3 | 82.329 |
| 12 | 22.15 | 5.1 | 0.6 | 1.4 | 90.971 |
| 13 | 23.45 | 3.9 | 1.1 | 0.6 | 92.751 |
| 14 | 25.45 | 4.2 | 1.2 | 0.8 | 92.141 |
| 15 | 12.45 | 4.3 | 1.8 | 1.1 | 76.707 |

The chart in Figure 5 compares the average number of segments filtered out in a session, the false positives and the false negatives. It can be observed that the mean of MFSC across the session is 3.49, whereas the mean of MFP and MFN are 1.3 and 0.9 respectively.

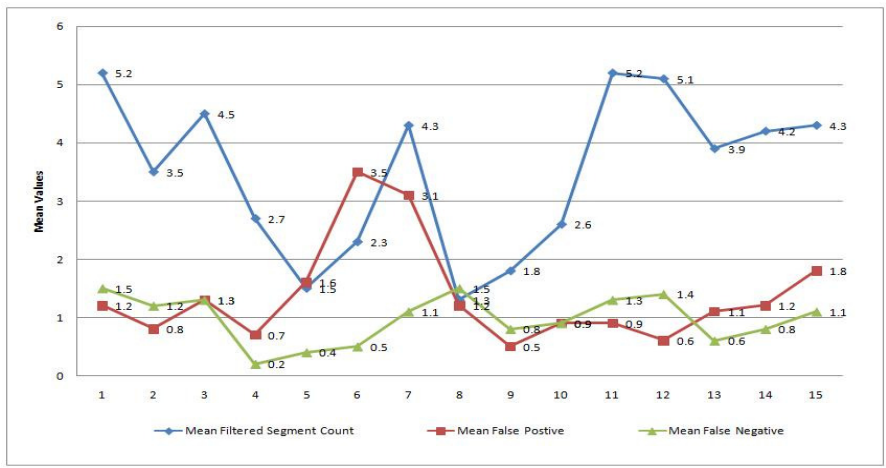

Figure 5. Comparison of MFSC, MFP and MFN

The chart in Figure 6 compares the Mean Segment Count with the accuracy. It can be observed that the mean accuracy of filtering across the session is 87.59 which confirm the efficiency of the proposed content filtering model.

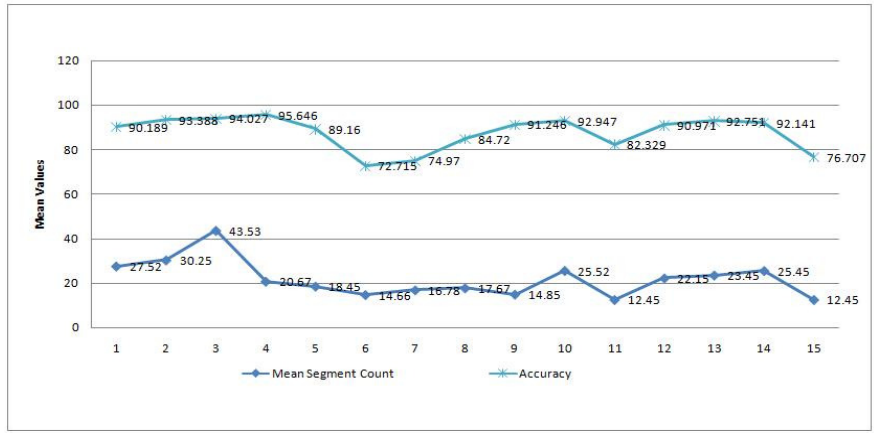

Figure 6. Comparison of MSC and Accuracy

## 5. CONCLUSIONS AND FUTURE DIRECTIONS

The proposed model for page filtering using segmentation and personalization renders the following advantages:

- Instead of blocking the entire page in cases where the content to be blocked is present only at a portion of the page, the proposed model provides a distinct benefit to user.

- Incorporation of personalization in the blocking process provides a tailor made content filtering system based on the user's needs.

The future directions for this research work are as listed below:

- In the proposed model the image filtering happens using the "alt" text provided with the image. In the future implementations some of the image analysis modules can be incorporated to make the image filtering much more efficient.

- Incorporation of the capability to handle languages other than English would make the system more efficient in the cases of non-English web pages.

**Authors**

K.S.Kuppusamy is an Assistant Professor at Department of Computer Science, School of Engineering and Technology, Pondicherry University, Pondicherry, India. He has obtained his Masters degree in Computer Science and Information Technology from Madurai Kamaraj University. He is currently pursuing his Ph.D in the field of Intelligent Information Management. His research interest includes Web Search Engines, Semantic Web. He has made 8 international publications.

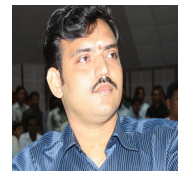

G. Aghila is a Professor at Department of Computer Science, School of Engineering and Technology, Pondicherry University, Pondicherry, India. She has got a total of 22 years of teaching experience. She has received her M.E (Computer Science and Engineering) and Ph.D. from Anna University, Chennai, India. She has published more than 55 research papers in web crawlers, ontology based information retrieval. She is currently a supervisor guiding 8 Ph.D. scholars. She was in receipt of Schrneiger award. She is an expert in ontology development. Her area of interest includes Intelligent Information Management, artificial intelligence, text mining and semantic web technologies.

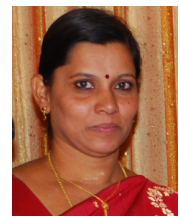